\documentclass[conference]{IEEEtran}
\IEEEoverridecommandlockouts
\usepackage{cite}
\usepackage{amsmath,amssymb,amsfonts}
\usepackage{graphicx}
\usepackage{textcomp}
\usepackage{xcolor}
\usepackage{outlines}
\usepackage{caption}
\usepackage{subcaption}
\usepackage[ruled,vlined]{algorithm2e} %linesnumbered
\usepackage{siunitx}
\usepackage[colorlinks=true,linkcolor=black,anchorcolor=black,citecolor=black,filecolor=black,menucolor=black,runcolor=black,urlcolor=black]{hyperref}
\usepackage{mathtools}
\usepackage{fancyhdr}
\usepackage[shortlabels]{enumitem}
\usepackage{float}
\usepackage{placeins}

\renewcommand{\footnoterule}{%
  \kern -3pt
  \hrule width 0.3\textwidth height 0.4pt
  \kern 2.6pt
}

\newcommand{\R}{\ensuremath{\mathbb{R}}}

\newcommand{\Probability}{\mathbb{P}}

\DeclareMathAlphabet\mathbfcal{OMS}{cmsy}{b}{n}

\newcommand{\E}{\mathbfcal{E}}
\newcommand{\Normal}{\mathcal{N}}

\newcommand{\mbf}[1]{{\mathbf{#1}}}
\newcommand{\transpose}{^\mathsf{T}}

\DeclareMathOperator*{\argmin}{arg\,min}

\newcommand{\norm}[1]{\left\Vert#1\right\Vert}

\newcommand{\regtext}[1]{\mathrm{\textnormal{#1}}}

\newcommand{\eye}{\mathbf{I}}
\newcommand{\zero}{\mathbf{0}}

\newcommand{\gps}{^\regtext{GPS}}
\newcommand{\icp}{^\regtext{ICP}}

\newcommand{\gt}{^\regtext{ref}}
\newcommand{\est}{^\regtext{est}}

\newcommand{\spoof}{_\regtext{spoof}}
\newcommand{\mean}{_\regtext{mean}}

\newcommand{\SO}{\mathbf{SO}}
\newcommand{\SE}{\mathbf{SE}}
\newcommand{\so}{\mathfrak{so}}
\newcommand{\se}{\mathfrak{se}}

\newcommand{\x}{\mathbf{x}}
\newcommand{\z}{\mathbf{z}}
\newcommand{\e}{\mathbf{e}}
\newcommand{\s}{\mathbf{s}}

\def\BibTeX{{\rm B\kern-.05em{\sc i\kern-.025em b}\kern-.08em
    T\kern-.1667em\lower.7ex\hbox{E}\kern-.125emX}}
\begin{document}

% Title
\title{Spoofing-Resilient LiDAR-GPS Factor Graph Localization with Chimera Authentication\\
% under Chimera Authentication?

% Identify applicable funding agency here. If none, delete this.
% \thanks{This work is sponsored by the Air Force Research Lab (AFRL) under grant number FA9453-20-1-0002.}
\thanks{The views expressed are those of the authors and do not reflect the official guidance or position of the United States Government, the Department of
Defense or of the United States Air Force. Statement from DoD: The appearance of external hyperlinks does not constitute endorsement by the United States
Department of Defense (DoD) of the linked websites, or the information, products, or services contained therein. The DoD does not exercise any editorial,
security, or other control over the information you may find at these locations.}

}

\author{\IEEEauthorblockN{Adam Dai}
\IEEEauthorblockA{\textit{Electrical Engineering} \\
\textit{Stanford University}\\
Stanford, USA \\
addai@stanford.edu}
\and
\IEEEauthorblockN{Tara Mina}
\IEEEauthorblockA{\textit{Electrical Engineering} \\
\textit{Stanford University}\\
Stanford, USA \\
tymina@stanford.edu}
\and
\IEEEauthorblockN{Ashwin Kanhere}
\IEEEauthorblockA{\textit{Aeronautics and Astronautics} \\
\textit{Stanford University}\\
Stanford, USA \\
akanhere@stanford.edu}
\and
\IEEEauthorblockN{Grace Gao}
\IEEEauthorblockA{\textit{Aeronautics and Astronautics} \\
\textit{Stanford University}\\
Stanford, USA \\
gracegao@stanford.edu}
}

\maketitle

\thispagestyle{fancy}

\begin{abstract}

% To protect against GPS spoofing attacks, the Air Force Research Lab (AFRL) has developed the Chips-Message Robust Authentication (Chimera) signal enhancement for the GPS L1C signal. 
% With a compatible receiver, users will be able to authenticate GPS measurements using the Chimera signal enhancement every 3 minutes if relying solely on Chimera GPS signals or every 1.5 or 6 seconds if accessing a secure out-of-band channel. 
% However, for fast-moving vehicles, the latency of even the faster channel (6 seconds) may still be too high, forcing users to decide between using high-latency, authenticated GPS measurements or low-latency, potentially vulnerable GPS measurements.

% Introduce AVs, sensor fusion/GPS with LiDAR
% Currently, all autonomous vehicles relying on GPS are vulnerable to spoofing attacks, which are capable of inducing errors in positioning that can severely compromise the safety of such vehicles.

Many vehicle platforms typically use sensors such as LiDAR or camera for locally-referenced navigation with GPS for globally-referenced navigation. 
% Experts predict that in the next two decades, autonomous vehicles will be prevalent on our roads, becoming a crucial part of our transportation infrastructure.
% These autonomous vehicles rely on sensors such as camera or LiDAR to navigate within their local surroundings, but to perform globally-referenced navigation, they must still rely on GPS when hioh-resolution globally-referenced maps are not available. 
However, due to the unencrypted nature of GPS signals, all civilian users are vulnerable to spoofing attacks, where a malicious spoofer broadcasts fabricated signals and causes the user to track a false position fix.
% The Air Force Research Lab (AFRL) has developed Chips-Message Robust Authentication (Chimera) to protect against GPS spoofing attacks---however, Chimera alone is not enough to ensure the safety of an autonomous vehicle. 
% For a standalone user with only access to the GPS signal, Chimera provides authentication every 3 minutes, whereas LiDAR-GPS sensor fusion typically requires a GPS update rate on the order of 1 Hz.
% The Air Force Research Lab has developed Chips-Message Robust Authentication (Chimera) to protect against such GPS spoofing attacks, but the Chimera authentication is not continuously available 
To protect against such GPS spoofing attacks, Chips-Message Robust Authentication (Chimera) has been developed and will be tested on the Navigation Technology Satellite 3 (NTS-3) satellite being launched later this year.
However, Chimera authentication is not continuously available 
%(only available once every 3 minutes for one of its service channels) 
and may not provide sufficient protection for vehicles which rely on more frequent GPS measurements. %(typically every second or faster).
In this paper, we propose a factor graph-based state estimation framework which integrates LiDAR and GPS while simultaneously detecting and mitigating spoofing attacks experienced between consecutive Chimera authentications.
% Our factor graph tightly-couples GPS pseudorange measurements with odometry measurements from LiDAR point cloud registration, and uses a chi-squared detector based on pseudorange residuals for spoofing detection.
Our proposed framework combines GPS pseudorange measurements with LiDAR odometry to provide a robust navigation solution. 
A chi-squared detector, based on pseudorange residuals, is used to detect and mitigate any potential GPS spoofing attacks.
We evaluate our method using real-world LiDAR data from the KITTI dataset and simulated GPS measurements, both nominal and with spoofing.
Across multiple trajectories and Monte Carlo runs, our method consistently achieves position errors under 5 \si{m} during nominal conditions, and successfully bounds positioning error to within odometry drift levels during spoofed conditions. 
% When to strategically use GPS during Chimera epoch, and limit the influence of spoofing.

% A naive algorithm with no spoofing detection/mitigation could have unbounded error for an unbounded spoofing attack. We are able to bound error to within lidar odometry drift
% TODO: add key numbers for results (e.g. "detects X% of spoofing attacks, etc.")

\end{abstract}

\section{Introduction}

% Localization is critical, sensor fusion and LiDAR-GPS pairing
% Cite Waymo, Cruise, as examples of AVs coming to the streets
Localization is a fundamental task for vehicle-related applications, such as autonomous driving or precision farming. Currently, state-of-the-art vehicle localization relies on sensor fusion, as various sensors possess different tradeoffs and advantages. Centimeter-level localization of self-driving cars has been demonstrated with fusion of LiDAR (Light Detection and Ranging), vision, and GPS, with evaluation on a real-world fleet of cars \cite{wan2018robust}. In particular, LiDAR and GPS have complementary advantages. LiDAR localization and odometry works well in structured environments, but struggles in empty areas lacking spatial features. Conversely, GPS struggles in structured environments due to signal blockage and multipath, but excels in open-sky conditions. 
% Many sensor fusion approaches leverage this strong LiDAR-GPS pairing \cite{..., ..., ...}

% GPS vulnerability to spoofing
However, GPS is vulnerable to spoofing, in which an attacker transmits fabricated GPS signals at higher power than the real signals, causing the victim to lock on to the fake signals. The attacker can then induce arbitrary errors to the victim’s GPS position estimate. For a vehicle running sensor fusion with GPS, these errors will propagate to the localization solution, compromising the safety of humans onboard or near the vehicle. Indeed, this vulnerability has been demonstrated in recent work, in which a well-designed GPS spoofing attack is able to cause an autonomous vehicle to crash with 97\% success rate \cite{shen2020drift}. 

% Chimera
As a countermeasure to GPS spoofing, the Air Force Research Lab (AFRL) has proposed the Chips-Message Robust Authentication (Chimera) signal enhancement for the GPS L1C signal \cite{anderson2017chips}. The Chimera signal enhancement punctures the L1C spreading code in the pilot channel with encrypted markers, which cannot be predicted beforehand, but can be verified via a digital signature provided to the user with a short latency. For standalone receivers, authentication is available every 3 minutes through the \textit{slow channel}, while users with access to secure Internet connection or an augmentation system can receive authentication every 1.5 or 6 seconds through the \textit{fast channel}. The time duration between consecutive Chimera authentications is referred to as the \textit{Chimera epoch}. 

% Chimera weaknesses
Nevertheless, for either the slow or fast channel, the Chimera authentication service is not continuously available. For applications such as self-driving, 1.5 seconds can easily make the difference between staying safe and crashing. Furthermore, for users relying on the slow channel, an attacker would have a large window of time to introduce spoofing errors. Prior works have addressed this issue through spoofing mitigation between Chimera authentications \cite{mina2022gps, kanhere2022factor}. These works use IMUs (inertial measurement units) and wheel encoders as trusted (i.e. unaffected by spoofing) sensors for fusion with GPS. However, the problem of spoofing detection and mitigation has yet to be explored for LiDAR-GPS sensor fusion.

\subsection{Contributions}

% Contributions after GG's review
In this work, we develop a novel spoofing detection and mitigation framework for LiDAR-GPS sensor fusion. This problem has received little attention in prior literature, and to the best of our knowledge, our solution is the first to examine the problem in the context of Chimera signal enhancement.
%incorporate benefits from the Chimera signal enhancement.
When combined with Chimera authentications, our approach mitigates the localization error induced by a spoofer over the Chimera epoch, which we experimentally validate using real LiDAR and simulated GPS measurements.

The key contributions of this work are:
\begin{enumerate}
    % \item We propose a spoofing-resilient factor graph optimization framework for state estimation using LiDAR odometry and GPS pseudoranges that detects and mitigates spoofing during a Chimera epoch.
    \item We perform tightly-coupled factor graph optimization with LiDAR odometry and GPS pseudoranges for accurate vehicle localization within the Chimera epoch.
    % \item During the Chimera epoch, we continuously monitor the unauthenticated GPS measurements using a chi-squared detector. When determined to be authentic, the GPS measurements are leveraged to minimize LiDAR odometry drift; otherwise, our framework places relies on the LiDAR odometry to mitigate effects of the spoofing attack.
    \item During the Chimera epoch, we use a chi-squared detector to determine the authenticity of GPS measurements. When measurements are deemed unauthentic, we mitigate the effects of the spoofing attack by relying on LiDAR odometry and excluding GPS.
    \item We validate our approach experimentally for the 3-minute Chimera slow channel, using real-world LiDAR measurements from the KITTI dataset and simulated GPS measurements. During nominal conditions, our approach maintains accuracy comparable to baseline methods. During spoofed conditions, our approach demonstrates consistent detection and mitigation of the attack across various trajectories and spoofing attacks. 
\end{enumerate}
To the best of our knowledge, we believe this is the first spoofing detection and mitigation approach for tightly-coupled GPS factor graph optimization.

\subsection{Paper Organization}

The remainder of this paper is organized as follows. Section~\ref{sec:related_work} surveys relevant literature to this work. Section~\ref{sec:preliminaries} introduces the problem statement and notation, and provides background on pose representation and factor graph optimization. Section~\ref{sec:approach} details our factor graph optimization and spoofing detection and mitigation framework. Section~\ref{sec:experiments} describes the setup and parameters for experimental validation, Section~\ref{sec:results} presents the experimental results, and Section~\ref{sec:conclusion} concludes this paper.
\section{Related Work} \label{sec:related_work}

%% -- Signposting
Our work bridges the areas of LiDAR-GPS sensor fusion and spoofing detection and mitigation in the context of Chimera GPS. In this section, we discuss existing approaches for LiDAR-GPS sensor fusion, followed by prior works addressing spoofing detection and resilient estimation.
% summary sentence

%% -- LiDAR-GPS sensor fusion
\subsection{LiDAR-GPS Sensor Fusion}
LiDAR-GPS sensor fusion approaches can be broadly separated in two main categories: \textit{filtering-based} and \textit{graph-based}.
% Filter-based
Filtering-based approaches rely on recursive Bayesian estimation as the underlying state estimation and fusion framework.
The most notable examples of Bayesian filters are the Kalman filter (KF), Extended Kalman Filter (EKF), and the Particle Filter (PF).
Several current state-of-the art LiDAR-GPS sensor fusion approaches rely on filtering~\cite{wan2018robust, li2021gil}.

% Graph-based
% Graph approaches becoming popular. In particular, tightly-coupled factor graphs with GPS range factors have been shown to be promising for sensor fusion
% (Graphs able to leverage large amount of sensor observations, process together in large windows)
% However, few spoofing detection/mitigation techniques have been developed for graph-based frameworks
% As a side plus, inherently (graph-based) allow for incorporating updated information 
Graph-based approaches encode vehicle states and sensor observations into a graph data structure, and employ graph optimization to solve for the optimal trajectory.
Over the past decade, there has been continually growing interest in factor graph optimization (FGO) for sensor fusion localization.
% Survey: filtering vs optimization approaches in SLAM
Recently, Wen et al.~\cite{wen2021factor} compared EKF and FGO localization approaches in a GPS challenged environment, and found that the FGO outperformed the EKF for the tightly-coupled case, in which GPS pseudoranges are incorporated directly into the graph.
The authors also showed that tightly-coupled FGO outperforms the loosely-coupled alternative, in which GPS position measurements are used as factors rather than pseudoranges.
%, rather than used to determine GPS position measurement (loosely-coupled).
Successful graph-based integrations of LiDAR and GPS have also been explored.
Chen et al.~\cite{chen2019probabilistic} present a Bayesian graph for fusion of LiDAR, GPS, and 3D building maps in order to localize a UAV in an urban environment.
The authors demonstrate significant improvement over a GPS-only Kalman filter approach, but the method relies on map matching with existing 3D building models to achieve accurate localization.
He et al.~\cite{he2022lidar} also leverage graph optimization to fuse LiDAR, IMU and GPS.
The authors evaluate their method on the KITTI dataset~\cite{geiger2013vision}, outperforming state-of-the-art LiDAR odometry approaches with meter-level accuracy, while also demonstrating their algorithm can be run in real-time at low latency.

% \red{Although, the idea of graph optimization is old, the technology is still seeing recent advancements - e.g. SymForce}

%% -- GPS spoofing
\subsection{Spoofing Detection and Resilient State Estimation}
However, none of the above LiDAR-GPS sensor fusion works address the vulnerability of GPS to spoofing attacks.
In 2020, Shen et al.~\cite{shen2020drift} demonstrated a spoofing method which is able to exploit the sensor fusion algorithm of~\cite{wan2018robust}, and induce large lateral deviations to the vehicle's state estimate, and consequently to the actual trajectory, during periods of low confidence.
With just 2 minutes of attack time, the spoofing algorithm is able to induce dangerous vehicle behavior with a 97\% success rate.
Outside of sensor fusion, GPS spoofing attack methods and detection strategies have received much attention~\cite{psiaki2016gnss}.
However, many detection strategies make assumptions about receiver capabilities or require additional functionality, such as multiple antennas.
% Chimera
Chimera is the first proposed authentication service for GPS signals and is set to be tested onboard the NTS-3 (Navigation Technology Satellite 3) platform scheduled for launch in 2023~\cite{hinks2021signal}.

%% -- Spoofing resilient GPS sensor fusion
Very recently, some works have begun to address the problem of spoofing-resilient GPS sensor fusion with Chimera.
% Cite Tara's SR Filter work too?
Mina et al.~\cite{mina2022gps} present a spoofing-resilient filter for continuous state estimation between Chimera authentications, which leverages IMU and wheel encoders as self-contained sensors to determine the trustworthiness of received GPS signals.   
Kanhere et al.~\cite{kanhere2022factor} use FGO to combine GPS, IMU and wheel odometry to perform spoofing mitigation with Chimera authentication.
The authors model the authentication state as switchable constraints~\cite{sunderhauf2012switchable} in the graph, and test their method on simulated trajectories using the fast channel authentication period of 6 seconds.

Our approach extends upon these prior works, integrating elements of the chi-squared detection scheme from prior works~\cite{tanil2017ins, mina2022gps} into a graph formulation.
Furthermore, while the factor graph approach of \cite{kanhere2022factor} only performs implicit mitigation, and is only evaluated for fast-channel application on a short (12 \si{s}) trajectory, our approach performs explicit detection and mitigation and is evaluated on trajectories spanning the slow-channel Chimera epoch of 3~minutes.
Additionally, our approach uses a tightly-coupled factor graph with GPS pseudorange factors, which has been found to outperform the loosely-coupled version in localization accuracy in prior work \cite{wen2021factor}.
Finally, we incorporate LiDAR as a new sensor in the realm of sensor fusion under Chimera.

%% -- Other papers to cite
%\red{Other papers to cite:} GPS/IMU Factor Graph~\cite{wen2021factor}, FGO Integrity Monitoring~\cite{wen2021integrity}, 
\section{Preliminaries} \label{sec:preliminaries}
% Preliminaries, Notation, Background

% TODO: add signposting
In this section, we present our problem statement and objective, discuss notation and models used in the paper, and cover relevant background on Lie groups and factor graph optimization.

\subsection{Problem Statement}
%% -- Problem Statement

We consider a vehicle equipped with a GPS receiver and LiDAR moving through an environment with continuous GPS availability.
During operation, the vehicle may be subject to GPS spoofing attacks which induce arbitrary bias error to the GPS measurements. 
However, the GPS receiver has access to slow channel Chimera authentication every 3 minutes.
Within the 3-minute Chimera epoch, our objective is to perform spoofing-resilient localization.
In particular, we wish to determine when to leverage the available, but not-yet-authenticated GPS measurements and when to fall back on the LiDAR measurements only.
In this way, we seek to improve localization performance when GPS is likely authentic, while remaining resilient to an experienced spoofing attack. %Our objective is to localize the vehicle, using its onboard GPS and LiDAR sensors, as accurately as possible across both nominal and spoofed conditions.
% Within the 3 minutes, our objective is to determine when to leverage GPS within the LiDAR-GPS graph framework to improve localization of the vehicle  
% Our objective is, within the 3 minute Chimera window, perform accurate localization under nominal (unspoofed) conditions, and determine when to leverage GPS within the LiDAR-GPS 

\subsection{Notation} \label{subsec:notation}
%% -- Notation

We model time as discrete, with $\Delta t$ denoting the discretization interval in seconds. The variable $k$ is used to denote the current time index, while the variable $i$ is used to denote an arbitrary time index.
% We denote the LiDAR sensor rate as $f\lidar$ and the GPS measurement rate $f\gps$ (both in \si{Hz}).
At time $k$, if a LiDAR measurement is available, we obtain a point cloud $\mbf{P}_k \in \R^{N_{\regtext{points}}\times 3}$ where $N_{\regtext{points}}$ is the number of points in the point cloud.
Likewise, if GPS is available at time $k$, we obtain a set of pseudorange measurements $\boldsymbol\rho_k = (\rho_k^{(1)}, \dots, \rho_k^{(m)}) \in \R^m$, where $\rho_k^{(j)}$ is the measured pseudorange to satellite $j$, and $m$ is the number of visible satellites.

% Chimera epoch length notation:
Recall that the time duration between successive authentications is referred to as the Chimera epoch.
The length of the Chimera epoch for the slow channel, in discretized timesteps, is then $N_{\regtext{epoch}} \coloneqq 180/\Delta t$.

% $\eye_{n\times n} \in \R^{n\times n}$ is used to denote $n\times n$ identity matrix, and $\zero_{n\times m} \in \R^{n\times m}$ the $n\times m$ matrix of zeros.
$\eye$ refers to the identity matrix, and $\zero$ to the matrix of zeros. 
Scalars are denoted with lowercase italics, vectors with lowercase boldface, and matrices with uppercase boldface.

\subsection{GPS Pseudorange Error Model} \label{subsec:pr_model}

We model the distribution of authentic, i.e. unspoofed, GPS pseudorange error as a zero-mean Gaussian $\Normal(0, \sigma\gps)$, where $\sigma\gps$ is the standard deviation of the pseudorange error.
Thus we can write
\begin{align} \label{eq:gps_model}
\rho_k = \bar{\rho}_k + \epsilon_k,\ \epsilon_k \sim \Normal(0, \sigma\gps),
\end{align}
where $\rho_k$ is the measured range and $\bar{\rho}_k$ is the true range.
We assume that clock bias effects have been removed from the measurements.

\subsection{$\SO(3)$ and $\SE(3)$ Lie groups} \label{subsec:lie_groups}

We present a brief background on the rotation and rigid body transformation Lie groups $\SO(3)$ and $\SE(3)$, as we use the corresponding 3D representations for the vehicle's 3D pose. 
% they are instrumental in representing the vehicle's 3D pose in manner conducive for factor graph optimization.
More extensive coverage of these topics can be found in \cite{blanco2021tutorial, barfoot2017state, sola2018micro}.
The vehicle's 3D pose at time $k$ is denoted as $\x_k \in \SE(3)$, where $\SE(3)$ is the Special Euclidean group of dimension 3. 
A pose in $\SE(3)$ consists of a rotation $\mbf{R} \in \SO(3)$, where $\SO(3)$ is the Special Orthogonal group of dimension 3, and a translation $\mbf{t} \in \R^3$. 
$\SO(3)$ is defined as 
$$\SO(3) = \{\mbf{R} \in \R^{3\times 3} \ | \ \mbf{R}\transpose \mbf{R} = \eye, \det \mbf{R} = 1\},$$
i.e. the set of all rotation matrices (orthogonal matrices with determinant 1).
$\SE(3)$ can then be represented as the Cartesian product of $\SO(3)$ with $\R^3$, i.e., $\SE(3) \sim \SO(3) \times \R^3$.
% SE(3) can be defined as the set of all [R t \\ 0 1] in R4x4 for which R in SO(3), t in R3
We can represent $\x_k$ with a transformation matrix 
$$\mbf{T}_k = \begin{bmatrix} \mbf{R}_k & \mbf{t}_k \\ \zero & 1 \end{bmatrix} \in \R^{4\times 4},$$ 
where $\mbf{R}_k \in \R^{3\times 3}$ is a rotation matrix representing orientation in the global frame, and $\mbf{t}_k \in \R^3$ is a translation vector representing position in the global frame.

The Lie groups $\SO(3)$ and $\SE(3)$ have associated Lie algebras denoted $\so(3)$ and $\se(3)$, with dimensionality 3 and 6 respectively. 
The Lie algebra can be thought of as the tangent space (linearization) of the manifold at the identity element, and are linear spaces upon which optimization may be done conveniently.
More precisely, there is an isomorphism from $\so(3)$ to $\R^3$, and from $\se(3)$ to $\R^6$, and any $\mbf{R} \in \SO(3)$ can be represented with a vector $\omega \in \R^3$, and similarly any $\mbf{T} \in \SE(3)$ can be represented with a vector $\nu \in \R^6$. 

The exponential map $\exp: \se(3) \mapsto \SE(3)$ maps from the tangent space $\se(3)$ to $\SE(3)$ (from $\nu$ to $\mbf{T}$), while the logarithmic map $\log: \SE(3) \mapsto \se(3)$ maps from $\SE(3)$ to its tangent space $\se(3)$ (from $\mbf{T}$ to $\nu$). 
Details on how the exponential and logarithmic map are computed can be found in \cite{blanco2021tutorial, barfoot2017state, sola2018micro}.

For $\mbf{x}, \mbf{y} \in \SE(3)$, the ``ominus" operator is defined as $\mbf{y} \ominus \mbf{x} = \log(\mbf{x}^{-1}\mbf{y}) \in \se(3)$ \cite{blanco2021tutorial}.
This operator allows us to compute the ``difference" of poses in $\SE(3)$ in linearized tangent space coordinates, and will be used later in defining the LiDAR odometry residual for our factor graph.

\subsection{Factor Graph Optimization} \label{subsec:fgo}

Our state estimation framework relies on factor graph optimization (FGO), in which a graph encoding vehicle poses and sensor measurements is optimized to determine the estimated trajectory.
In this section, we provide a brief background on general factor graph formulation. 
More details can be found in \cite{dellaert2017factor} and \cite{grisetti2010tutorial}.

A factor graph consists of a set of nodes $\x = \{\x_1,\dots,\x_N\}$ which represent poses or states, and a set of edges or factors $\E$ which represent sensor measurements which constrain the graph.
A sensor observation linking nodes $\x_i$ and $\x_j$ is denoted $\z_{i,j}$ with associated information matrix $\Omega_{i,j}$, which is defined as the squared inverse of the measurement covariance matrix: $\Omega_{i,j} \coloneqq \Sigma_{i,j}^{-1}$.
Each sensor has an associated measurement model $\hat{\z}_{i,j}(\x_i, \x_j)$, which is used to define a residual $\e_{i,j}(\x_i, \x_j) = \z_{i,j} - \hat{\z}_{i,j}(\x_i, \x_j)$ for each factor.

Optimizing the factor graph consists of minimizing the following objective
\begin{align}
    \mbf{F}(\x) = \sum_{(i,j)\in\E} \e_{ij}\transpose \Omega_{ij} \e_{ij}
\end{align}
which is the sum of information-normalized squared error of the residuals.
This objective represents the negative log-likelihood of the vehicle poses given the sensor measurements.
Thus, solving the optimization problem 
\begin{align}
    \x^* = \argmin_{\x} \mbf{F}(\x).
\end{align}
yields the optimal set of poses $\x^*$ given our measurements.

The optimization is done by linearizing $\mathbf{F}$ and iteratively solving for updates to the state $\x$. 
For each edge $(i,j)\in\E$, the gradient $\mbf{b}_{ij}$ and Hessian $\mbf{H}_{ij}$ are computed as
\begin{align}
    \mbf{b}_{ij} = \e_{ij}\transpose \Omega_{ij} \mbf{J}_{ij}, \\
    \mbf{H}_{ij} = \mbf{J}_{ij}\transpose \Omega_{ij} \mbf{J}_{ij}, 
\end{align}
where $\mbf{J}_{ij}$ is the Jacobian of $\e_{ij}(\x)$.
The individual gradients and Hessians are then accumulated to form the gradient and Hessian for the entire graph, $\mbf{b} = \sum \mbf{b}_{ij}$ and $\mbf{H} = \sum \mbf{H}_{ij}$, and the linear system
\begin{align}
\mbf{H} \Delta \x^{*} = -\mbf{b}
\end{align}
is solved with sparse Cholesky factorization to find the optimal update $\Delta \x^{*}$, which is applied to the state $\x \coloneqq \x + \Delta \x^{*}$.

\section{Approach} \label{sec:approach}

\begin{figure}
    \centering
    \includegraphics[width=0.47\textwidth]{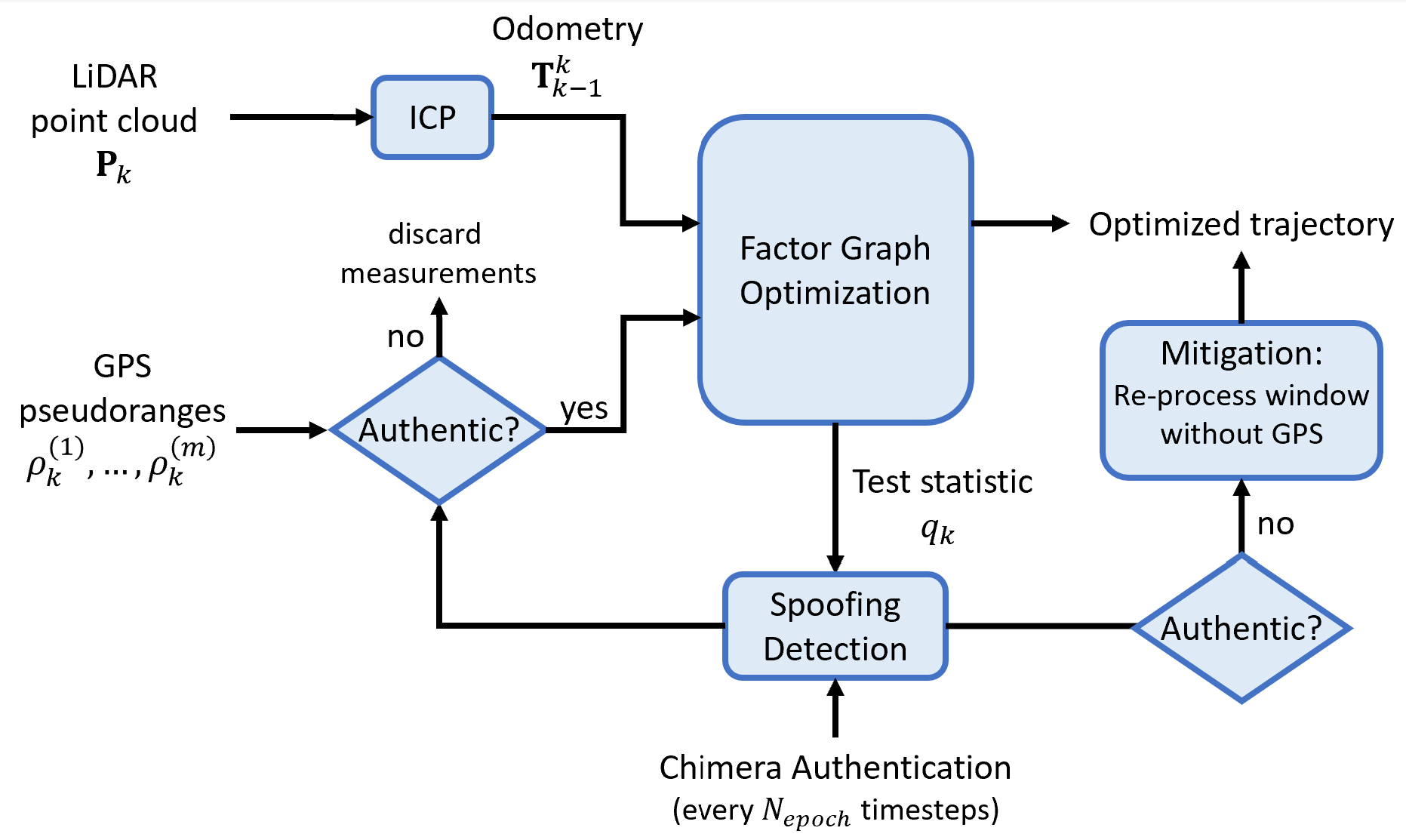}
    \caption{Block diagram illustrating our spoofing-resilient LiDAR-GPS factor graph framework. We operate over a window of measurements, for which LiDAR point clouds are registered with ICP to produce odometry measurements and combined with GPS pseudorange measurements in the factor graph. The vehicle trajectory over the window is estimated via factor graph optimization, and a chi-squared test statistic is computed from GPS residuals. If the statistic exceeds a threshold, an attack is detected, and it is mitigated by re-processing the window without GPS, and future GPS measurements are discarded from the factor graph. Chimera authentication overrides the detector as spoofing decision truth when available. }
    \label{fig:block_diagram}
\end{figure}

We now describe the details of our spoofing-resilient LiDAR-GPS factor graph approach.
Fig.~\ref{fig:block_diagram} shows a high-level block diagram of the framework.

\subsection{LiDAR-GPS Factor Graph}

\begin{figure}
    \centering
    \includegraphics[width=0.47\textwidth]{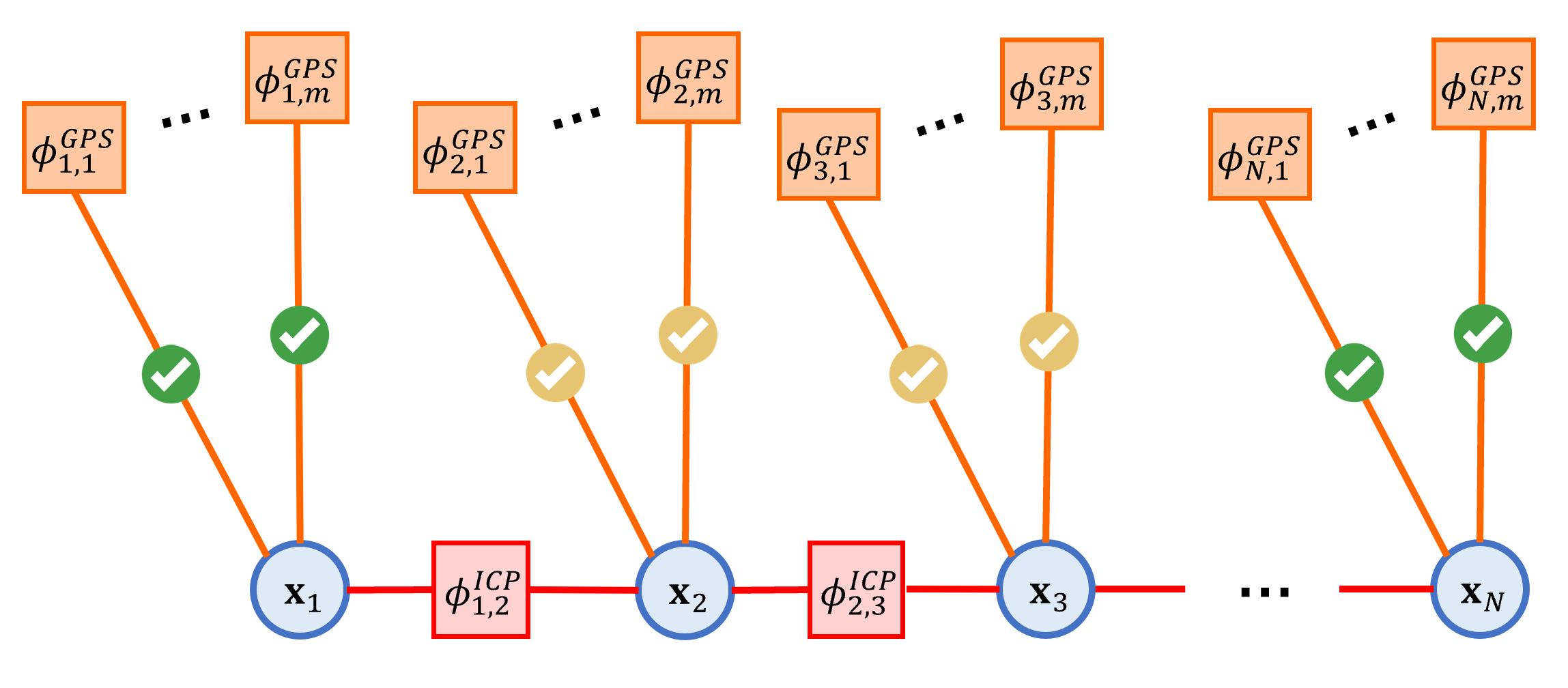}
    \caption{LiDAR-GPS factor graph structure over the Chimera epoch. Consecutive nodes are linked by LiDAR odometry factors $\phi\icp_{i,i+1}$, while GPS range factors $\phi\gps_{i,j}$ (where $j$ is satellite index) independently connect to nodes. The check symbols indicate that GPS has been deemed authentic, either by our detector (yellow) or by Chimera authentication (green)---in the unauthentic case GPS is excluded from the factor graph.}
    \label{fig:factor_graph}
\end{figure}

Our approach revolves around maintaining a tightly-coupled factor graph which integrates LiDAR and GPS for both localization and spoofing detection and mitigation, the structure of which is shown in Fig.~\ref{fig:factor_graph}.
The nodes of our factor graph are vehicle poses $\x_i \in \SE(3)$ as described in Section~\ref{subsec:lie_groups}.
The measurement models (Equations~\ref{eq:fgo_gps_model} and \ref{eq:fgo_icp_model}) and residuals (Equations~\ref{eq:fgo_gps_residual} and \ref{eq:fgo_icp_residual}) for GPS pseudoranges and LiDAR odometry factors are detailed in the following subsections.

% Loosely coupled

\subsubsection{GPS Pseudorange Factors}
% Tightly coupled
Given pose $\x_i$ with position component $\mbf{t}_i$ and satellite $j$ at position $\s_i^{(j)}$ at time $i$, the expected GPS pseudorange measurement from satellite $j$ to node $i$ is
\begin{align} \label{eq:fgo_gps_model}
    \hat{\rho}_i^{(j)}(\x_i) = \norm{\mbf{t}_i - \s_i^{(j)}}_2.
\end{align}
Now, given received pseudorange measurement $\rho_i^{(j)}$, the GPS residual function can be defined as
\begin{align} \label{eq:fgo_gps_residual}
    \e_{i,j}\gps(\x_i) = \rho_i^{(j)} - \hat{\rho}_i^{(j)}(\x_i).
\end{align}
As the residual is a scalar in this case, the information matrix is also a scalar, $(\sigma\gps)^{-2}$, where $\sigma\gps$ is the standard deviation of authentic GPS pseudoranges from Section~\ref{subsec:pr_model}.
In this work, we do not consider the effect of clock bias states in our pseudorange factors and simulated pseudorange measurements, but we will incorporate this into our measurement model in our future works.

% A constant information matrix $\Omega\gps$ is used for GPS measurements, and is computed as $\Omega\gps = (Q\gps)^{-1}$, where $Q\gps$ is the covariance of authentic GPS measurements from \eqref{eq:gps_model}. 

\subsubsection{LiDAR Odometry Factors}
For LiDAR, we first use point-to-plane ICP (Iterative Closest Point)~\cite{besl1992method} to register successive point clouds and produce an odometry measurement. 
Given point clouds $\mbf{P}_i$ and $\mbf{P}_{i+1}$ at times $i$ and $i+1$, ICP produces an odometry measurement $\mbf{T}_i^{i+1}$:
\begin{align} \label{eq:icp}
    \mbf{T}_i^{i+1} \coloneqq \regtext{icp}(\mbf{P}_i, \mbf{P}_{i+1}).
\end{align}
$\mbf{T}_i^{i+1}$ is a rigid body transformation in $\SE(3)$, and can be written as
\begin{align} 
    \mbf{T}_i^{i+1} = \begin{bmatrix} \mbf{R}_i^{i+1} & \mbf{t}_i^{i+1} \\ \zero & 1 \end{bmatrix} \in \R^{4\times 4},
\end{align}
where $\mbf{R}_i^{i+1} \in \R^{3\times 3}$ is the relative rotation between poses at times $i$ and $i+1$ and $\mbf{t}_i^{i+1} \in \R^3$ is the relative translation between poses at times $i$ and $i+1$.
The LiDAR odometry measurement model is the expected transformation between poses $i$ and $i+1$:
\begin{align} \label{eq:fgo_icp_model}
    \hat{\mbf{T}}_{i,i+1}(\x_i, \x_{i+1}) = \x_{i+1} (\x_i)^{-1}.
\end{align}
The LiDAR odometry residual function is then defined as 
\begin{align} \label{eq:fgo_icp_residual}
    \e_{i,i+1}\icp(\x_i,\x_{i+1}) = \mbf{T}_i^{i+1} \ominus \hat{\mbf{T}}_{i,i+1},
\end{align}
where $\ominus$ is the ``ominus" operator defined in Section~\ref{subsec:lie_groups}. 
Intuitively, this residual is the difference between the expected and measured odometry transformation in $\se(3)$ tangent space coordinates.
Thus, $\e_{i,i+1}\icp(\x_i,\x_{i+1}) \in \R^6$, and the information matrix for LiDAR odometry measurements is denoted by $\Omega\icp \in \R^{6\times 6}$.

\subsubsection{Optimization}
We optimize the graph in a sliding window fashion.
The window size is denoted by $N$, and as new nodes and measurements are added to the graph, previous nodes and edges are removed in order to maintain the maximum number of nodes in the graph as $N$.
% The window size is denoted by $N$, with the current window up to time index $k$ containing nodes $\{\x_i\},\ i \in [k-N+1, k]$.
% At each iteration, the window is shifted $N\shift$ timesteps to obtain a new window $[N\shift + k-N+1, N\shift + k]$.
The objective for our factor graph over a single window can thus be written as
\begin{align}
    \mbf{F}(\x) = 
    \sum_{i=1}^{N} \sum_{j=1}^m (\e_{i,j}\gps)\transpose (\sigma\gps)^{-2} \e_{i,j}\gps \\
    + \sum_{i=1}^{N-1} (\e_{i,i+1}\icp)\transpose \Omega\icp \e_{i,i+1}\icp
\end{align}
%, where $m$ is the number of visible satellites during the window.
The optimization carried out as described in Section~\ref{subsec:fgo}.

\subsection{Spoofing Detection and Mitigation} \label{subsec:spoofing_mitigation}

To perform detection between Chimera authentication times, we design a chi-squared spoofing detector within our factor graph framework.
Our detector computes a test statistic $q_k$ at time $k$ based on the information-normalized residuals of the GPS factors over the current window:
% \begin{align}
% q_k = \sum_{i=k-N\window+1}^k (\e_i\gps)\transpose \Omega_i\gps \e_i\gps
% \end{align}
\begin{align}
q_k = \sum_{i=i}^N \sum_{j=1}^m (\e_{i,j}\gps)\transpose (\sigma\gps)^{-2} \e_{i,j}\gps
\end{align}
Note that we do not include a normalization term based on the state estimate uncertainty (as typically done in the chi-squared detector with Kalman filter) as the factor graph does not maintain an estimate of state uncertainty.
We then compare $q_k$ with a threshold $\tau$, which is pre-computed based on user-specified false alarm requirements.
If $q_k > \tau$, then it is determined that spoofing is present in the measurements, otherwise the measurements are deemed authentic.

We now derive the computation of the threshold $\tau$.
When the received GPS measurements are authentic and follow the nominal distribution with zero-mean Gaussian noise as shown in Equation~\eqref{eq:gps_model}, the GPS residuals $\e_{i,j}\gps$ are distributed according to $\Normal(0, \sigma\gps)$ (assuming the estimated positions from FGO are close to ground-truth).
Then, since the test statistic $q_k$ is computed from squaring the residuals (of which there are $Nm$) and normalizing by $(\sigma\gps)^{-2}$, $q_k$ follows a central chi-squared distribution with $n = Nm$ degrees of freedom.
Thus, given a desired false alarm rate $\alpha$ to remain under, i.e., $\Probability(\regtext{detection} \, | \, \regtext{not spoofed}) \le \alpha$, we desire $\Probability(q_k \le \tau) = 1 - \alpha$ for nominal conditions.
Therefore, we compute $\tau$ as
\begin{align}
\tau = \Phi^{-1}(1-\alpha;\ n=Nm)
\end{align}
where $\Phi^{-1}$ is the inverse cumulative distribution function (CDF) of the chi-squared distribution with $n=Nm$ degrees of freedom. 

%\subsection{Spoofing Mitigation} 

If spoofing is detected at time $k$, i.e., $q_k > \tau$, then any future GPS measurements are deemed unauthentic and the FGO henceforth proceeds with LiDAR only.
Additionally, the current window is also re-processed with LiDAR measurements only, and GPS measurements discarded.

% all the most recent GPS measurements up until the previous Chimera authentication are assumed to be potentially unauthentic.
% We then remove the GPS edges in the current Chimera epoch from the graph by setting their information values to 0, effectively removing their influence on the graph optimization.
% The graph optimization is then re-run to form a new estimate which excludes the potentially spoofed measurements.

\subsection{Chimera Authentication}

After $N_{\regtext{epoch}}$ timesteps have passed, we receive Chimera authentication, which indicates whether the GPS measurements in the past Chimera epoch are authentic or unauthentic.
If the Chimera authentication determines the GPS measurements to be authentic, we leverage this information and rely on the received measurements within our factor graph for $N$ measurements, which corresponds to the window size of our factor graph. %disable our detector for a short time window to eliminate the risk of a false alarm.
%
%Additionally, based on the expected error or drift of an odometry-based approach, one may choose to re-initialize the trajectory estimate at the new authenticated GPS position.
However, if authentication fails, then we perform the same mitigation steps outlined in Section~\ref{subsec:spoofing_mitigation}, where we discard GPS measurements and rely only on the LiDAR sensor. 
At this point, the spoofed victim could discontinue nominal operations and proceed according a fail-safe protocol, such as safely pulling over to the side of road, the specifics of which are outside the scope of this work.
% Could say that when we get authentication, we fix the estimated 
\section{Experiments} \label{sec:experiments}

% TODO: add signposting
We now describe details of our experimental validation, including the dataset used, spoofing attacks considered, baselines which we compare to, and parameters choices for our implementation.

\subsection{KITTI Dataset}
% Describe dataset
We evaluate our approach using LiDAR data from the KITTI dataset \cite{geiger2013vision} and simulated GPS pseudorange measurements based on the ground-truth positions and satellite ephemeris.
In order to test our algorithm's ability to detect and mitigate spoofing attacks over a 3-minute slow-channel Chimera epoch, we select all sequences from the raw data recordings of duration 3 minutes or longer.
Table~\ref{tab:kitti_sequences} lists the four sequences used in our experiments, their total duration in seconds, and the abbreviations used to refer to each one throughout the remainder of the paper.
\begin{table}[htbp]
\caption{KITTI Sequences}
    \begin{center}
    \begin{tabular}{|c|c|c|}
        \hline
        Sequence name & Abbreviation & Total duration \\
        \hline
        2011\_09\_30\_drive\_0018 & 0018 & 276 \si{s} \\
        2011\_10\_03\_drive\_0027 & 0027 & 455 \si{s} \\
        2011\_09\_30\_drive\_0028 & 0028 & 518 \si{s} \\
        2011\_10\_03\_drive\_0034 & 0034 & 466 \si{s} \\
        \hline
    \end{tabular}
    \label{tab:kitti_sequences}
    \end{center}
\end{table}

% 0018 - 4:36 min
% 0028 - 8:38 min (6 min when starting at 1550)
% 0027 - 7:35 min
% 0034 - 7:46 min
For all sequences, we consider a 200 second segment of the trajectory, which contains a full 180 second Chimera slow channel epoch.
We simulate the Chimera authentication as occurring successfully at the first time step of the trajectory.
As a result, the second Chimera authentication time occurs 180 seconds into the trajectory, and we simulate this as a failed authentication for each of the simulated GPS spoofing test scenarios described in Section~\ref{subsec:spoofing_attacks}.

For all tested trajectories, we use the synced and rectified data in order to handle LiDAR motion distortion effects.
To simulate GPS pseudorange measurements, we compute GPS satellite positions over time $\s_i^{(j)}$ using ephemeris data pulled from the location and timestamps of each sequence and follow the measurement model outlined in Section~\ref{subsec:pr_model}.
% Visibility and DOP info: between 9 to 10 satellites visible, DOP around 1.5 for each sequence
% The number of visible satellites for each sequence varies between 9 and 10, with average position dilution of precision (PDOP) values ranging between 1.52 to 1.97.
For ground-truth reference trajectories, we use the OXTS ground-truth positions and orientations provided by KITTI.

\subsection{Spoofing Attacks} \label{subsec:spoofing_attacks}
%% -- Spoofing description

We simulate spoofing attacks on the vehicle by generating a spoofed reference trajectory with added biases, then computing spoofed GPS pseudoranges based on the spoofed reference trajectory.
Specifically, we consider a ramping attack that begins between the Chimera authentications, in which the spoofer introduces a bias which starts small and steadily ramps up to a large error.
This type of attack is typically the most difficult to detect, as the spoofer can gradually induce error without any sudden jumps to alert a standard RAIM solution.
And although the bias may start small, a spoofing victim under this attack can still experience significant positioning error over a sufficient time window, such as the 3 minute slow channel Chimera epoch.

\begin{figure*}
     \centering
     \begin{subfigure}[b]{0.49\textwidth}
         \centering
         \includegraphics[width=\textwidth]{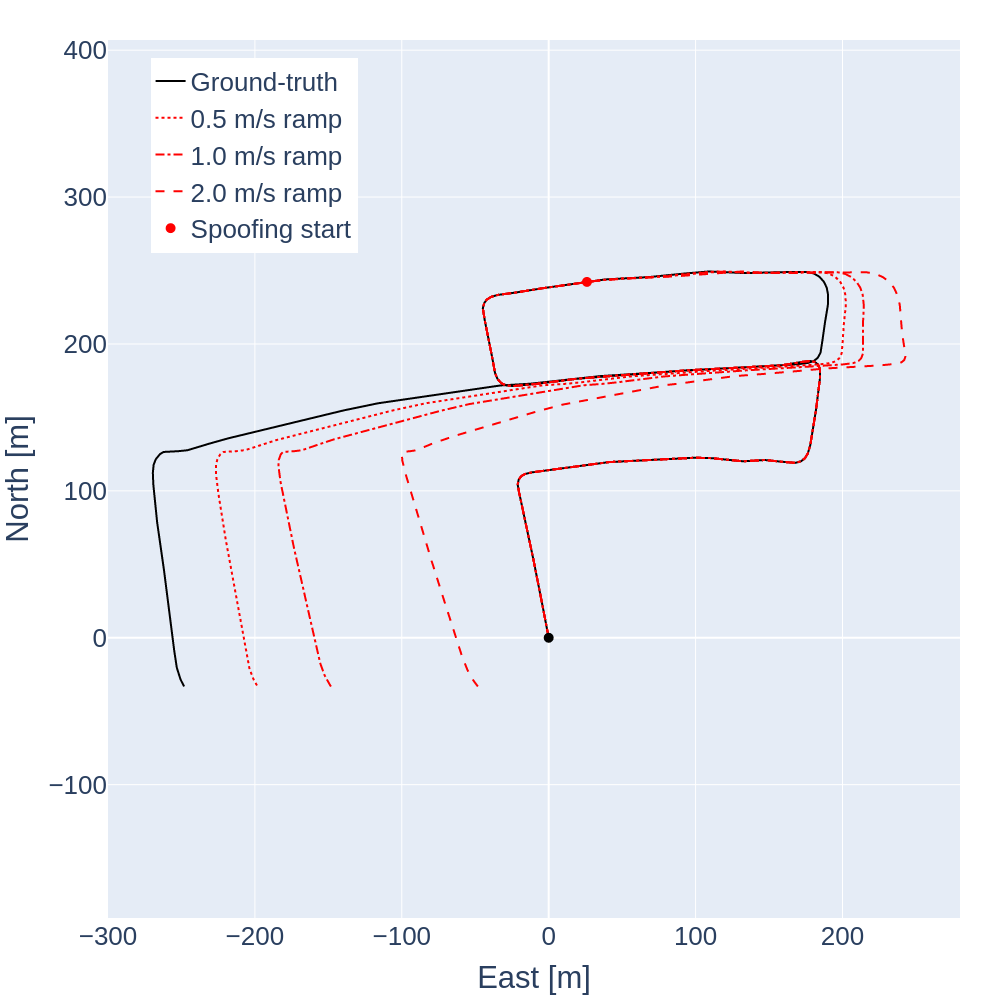}
         \caption{0018 reference trajectory with spoofed trajectories.}
         \label{fig:0018_traj}
     \end{subfigure}
     \hfill
     \begin{subfigure}[b]{0.49\textwidth}
         \centering
         \includegraphics[width=\textwidth]{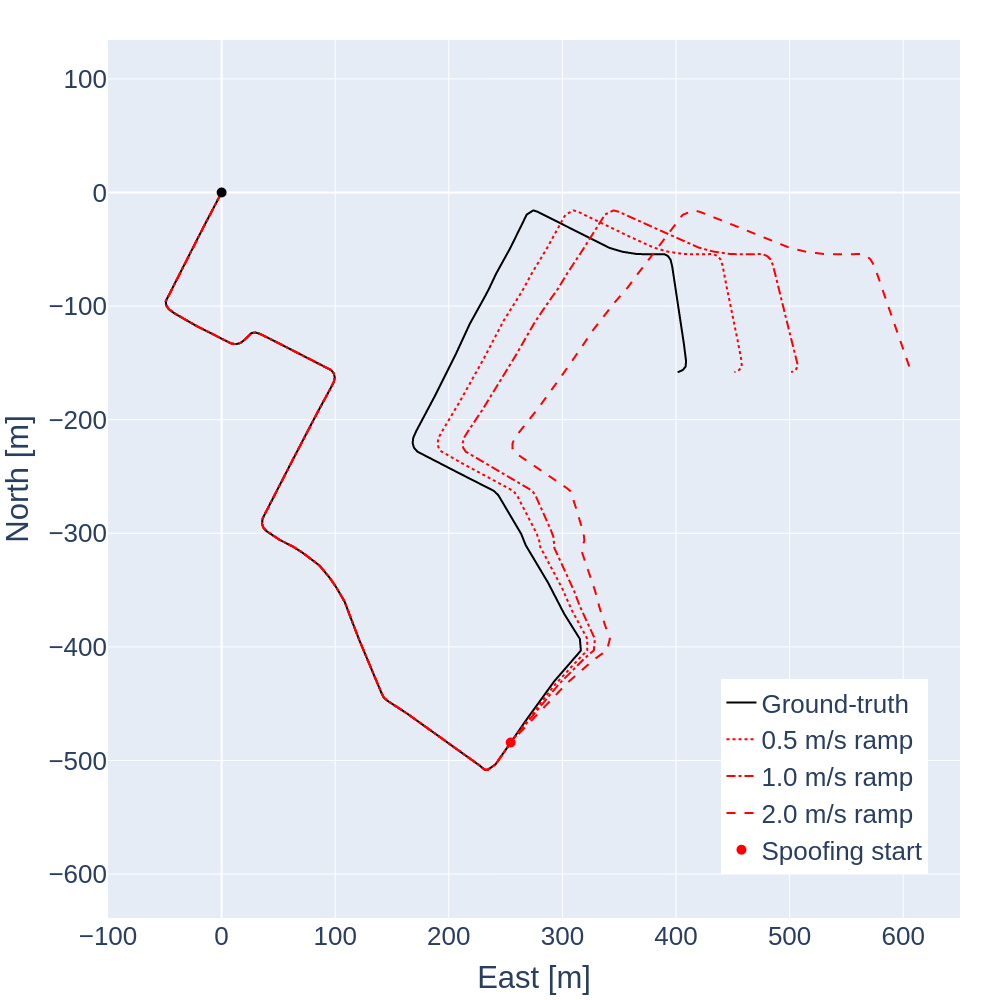}
         \caption{0027 reference trajectory with spoofed trajectories.}
         \label{fig:0027_traj}
     \end{subfigure}
     \hfill
     \begin{subfigure}[b]{0.49\textwidth}
         \centering
         \includegraphics[width=\textwidth]{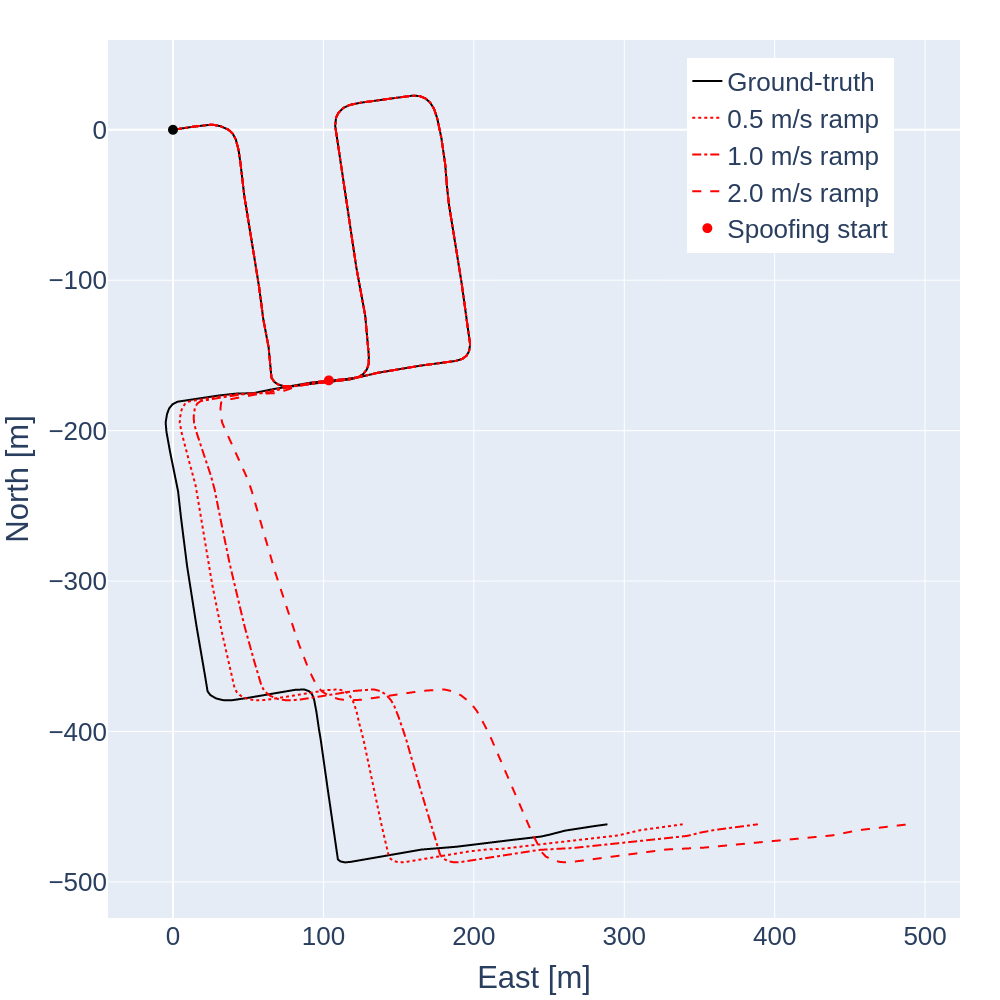}
         \caption{0028 reference trajectory with spoofed trajectories.}
         \label{fig:0028_traj}
     \end{subfigure}
     \hfill
     \begin{subfigure}[b]{0.49\textwidth}
         \centering
         \includegraphics[width=\textwidth]{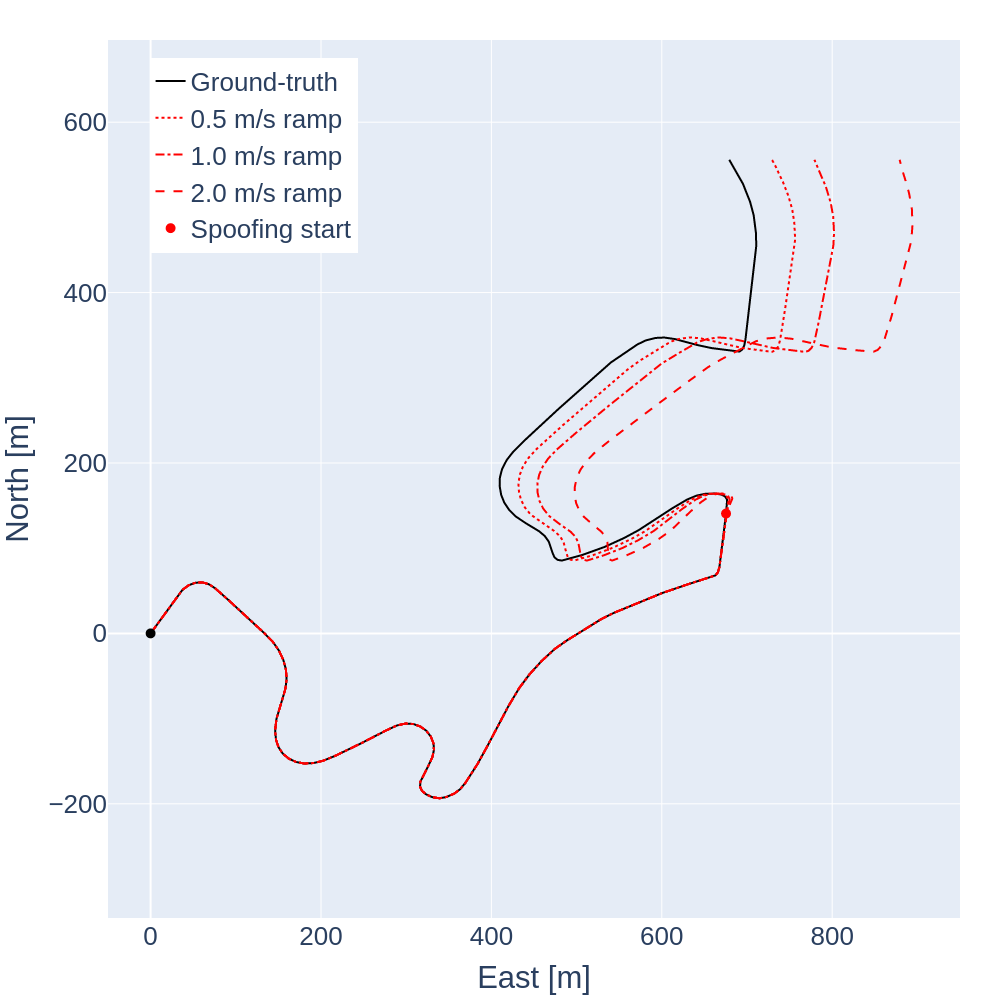}
         \caption{0034 reference trajectory with spoofed trajectories.}
         \label{fig:0034_traj}
     \end{subfigure}
    \caption{KITTI ground-truth reference trajectories shown with spoofed reference trajectories. The spoofing attacks begin at 100 seconds, introducing ramping error in the ENU positive $x$ (East) direction. Attacks for three different ramping rates are shown in different styles of red dashed lines.}
    \label{fig:spoof_trajs}
\end{figure*}

For our experiments, we use a ramping bias which starts at 0 \si{m}, and begins linearly increasing at rate $r\spoof$ \si{m/s} from time $T\spoof$ onward.
We add the spoofing bias to the ENU positive $x$ (East) direction, and choose $T\spoof = 100$ \si{s} for a total spoofing duration of 100 seconds.
% (It is assumed that a spoofer would require time to initiate spoofing after the first Chimera authentication, and thus a early attack stack is unrealistic)
We run experiments for ramp rates of $r\spoof = $ 0.5 \si{m/s}, 1.0 \si{m/s}, and 2.0 \si{m/s}, for maximum bias of 50 \si{m}, 100 \si{m}, and 200 \si{m} respectively.
Fig.~\ref{fig:spoof_trajs} shows the reference trajectories for each sequence, and spoofed trajectories for the chosen ramp rates.

% Spoofed trajectories

\subsection{Metrics and Baselines for Comparison} \label{subsec:baselines}
%% -- Baselines 
%  - Pure lidar odometry
%  - Blind FGO
%  - Spoofed FGO
In our experiments we compare our approach with two baseline approaches. 
The first baseline is ``Odometry only," in which only LiDAR odometry is used to localize the vehicle between Chimera authentications, and GPS measurements are only used at the slow channel 3 minute interval.
The second baseline is ``Naive FGO," in which LiDAR-GPS factor graph optimization produces a fused state estimate but no spoofing detection or mitigation is employed.
Finally, ``SR FGO" refers to our spoofing-resilient LiDAR-GPS factor graph optimization approach presented in Section~\ref{sec:approach}.

For characterizing performance, we consider $L^2$ norm position error, which is calculated as 
$e_k = \norm{\mbf{t}_k\gt - \mbf{t}_k\est}_2$ for each time index $k$ of the trajectory, where $\mbf{t}_k\gt$ is the reference trajectory position at time $k$ and $\mbf{t}_k\est$ is the estimated trajectory position at time $k$.
In addition, we consider two metrics: mean $L^2$ norm position error and maximum $L^2$ norm position error, which are simply computed as 
$e\mean = \regtext{mean}_k(e_k)$
and
$e_\regtext{max} = \regtext{max}_k(e_k)$
respectively, and hereafter referred to more concisely as mean error and max error.

\subsection{Parameters}
% Parameters
We discretize time with $\Delta t = 0.1$ \si{s}, and use LiDAR point clouds from the KITTI sequences taken at 10 \si{Hz}.
We use $\Sigma\icp = \regtext{diag}(0.01, 0.01, 0.01, 0.05, 0.05, 0.05)$ as the standard deviation of LiDAR ICP odometry measurements, where the 0.01 values correspond to the rotational components of $\se(3)$ (equivalent to 0.01 \si{rad} std.) and the 0.05 values correspond to the translational components (0.05 \si{m} std.). 
We simulate GPS measurements at 1 \si{Hz} and take $\sigma\gps = 7.0$ \si{m} according to the typical User-equivalent Range Error (UERE) for a single-frequency receiver \cite{kaplan2017understanding}.
We choose a window size of $N = 100$, and shift the window by 10 steps per iteration.
$\alpha=0.001$ is chosen as the false alarm rate for our detector.

For point-to-plane ICP LiDAR registration, we use the Open3D \cite{zhou2018open3d} function \texttt{registration\_icp} with parameter \texttt{TransformationEstimationPointToPlane()} class and threshold value of 1.0.
We use SymForce \cite{Martiros-RSS-22}, a recently developed state-of-the-art symbolic computation library for robotics applications, as the factor graph optimization backend for our method.
% We leverage the SymForce \texttt{Factor} and \texttt{Optimizer} classes, 
% More details can be found in our code available online at our GitHub repository\footnote{\url{https://github.com/Stanford-NavLab/chimera_fgo}}.
Our code is available online at our GitHub repository\footnote{\url{https://github.com/Stanford-NavLab/chimera_fgo}}.

\begin{figure*}[t]
     \centering
     \begin{subfigure}[b]{\textwidth}
         \centering
         \includegraphics[width=\textwidth]{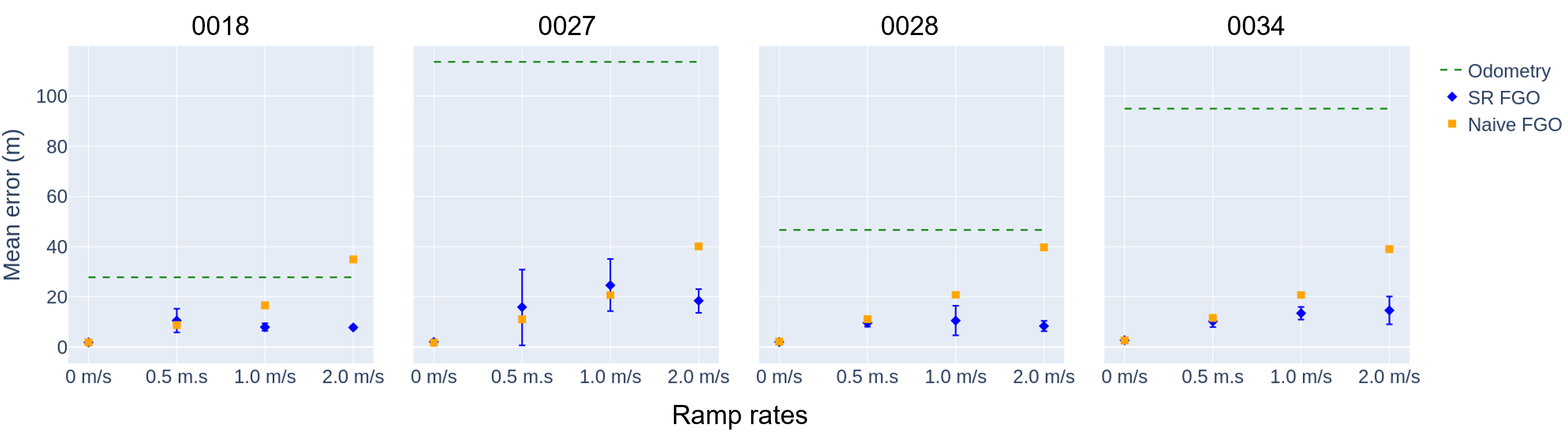}
         \caption{Mean $L^2$ norm position error compared across increasing ramp rates.}
         \label{fig:mean_scatter}
     \end{subfigure}
     \hfill
     \vspace{2mm}
     \begin{subfigure}[b]{\textwidth}
         \centering
         \includegraphics[width=\textwidth]{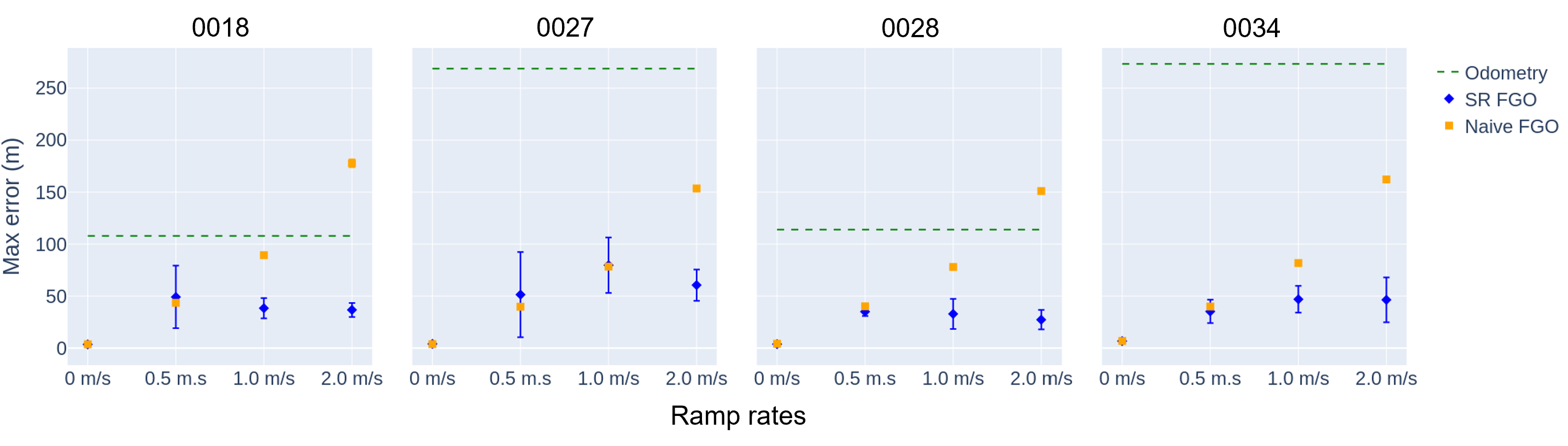}
         \caption{Max $L^2$ norm position error compared across increasing ramp rates.}
         \label{fig:max_scatter}
     \end{subfigure}
    \caption{Comparison of our approach with baselines in mean and maximum trajectory error for different sequences and spoofing ramp rates. Each subplot shows the mean/maximum trajectory error values of LiDAR odometry (green), Naive FGO (red), and SR FGO (blue) across increasing spoofing attack rates for a specific sequence. LiDAR odometry is unaffected by spoofing, and is thus shown as horizontal lines, while results for Naive FGO and SR FGO are each averaged over 10 Monte Carlo runs, with error bars indicating one standard deviation. Compared to the Naive FGO, SR FGO mitigates errors as spoofing ramp rate increases.}
    \label{fig:scatter_comparison}
\end{figure*}
\section{Results} \label{sec:results}

% TODO: add signposting
Now we present the experimental validation results of our spoofing-resilient factor graph algorithm. 
We run our algorithm along with the two baselines (Section~\ref{subsec:baselines}) on four KITTI sequences (Table~\ref{tab:kitti_sequences}), for both nominal GPS measurements and various ramping GPS spoofing attacks (shown in Fig.~\ref{fig:spoof_trajs}). 
We compare performance in terms of $L^2$ norm error over time, as well as mean and max $L^2$ norm error, and also include case studies on window size variation and detection statistics.

\subsection{Comprehensive Comparison}
%& -- Show comparison to baselines across all sequences and attack sizes

We begin by presenting a comprehensive comparison of our approach against the two baselines considered, across the different KITTI sequences and multiple spoofing ramp rates.
These results are illustrated in Fig.~\ref{fig:scatter_comparison}.
%as well as shown in Table~\ref{}.
We see that for all sequences, the mean and max $L^2$ norm error of our SR FGO approach remains under that of LiDAR odometry only.
In particular, as the spoofing attack rate increases, the Naive FGO mean and max errors increase, and in some cases eventually exceed the levels of odometry drift, whereas the SR FGO errors are successfully mitigated in all cases and remain bounded under odometry.

% For the case of Sequence 0027, we have observed that the LiDAR odometry drift direction paired with the spoofing attack direction induces a significant heading error which causes the SR FGO method to accumulate large errors after detection and switching to odometry only.
% Thus, the effectiveness of our approach does depend on the nature of the trajectory and LiDAR data. 
% However, we have found that this issue can be addressed by increasing the window size, and thus also increasing the ``lookahead" time for detection as well as re-processing window, as is shown in the later results of Section~\ref{subsec:window_comp}.

\subsection{Comparison of Errors over Time under Spoofed Conditions}
%& -- Show in-depth results for spoofed case of 200 m

\begin{figure*}
     \centering
     \begin{subfigure}[b]{0.49\textwidth}
         \centering
         \includegraphics[width=\textwidth]{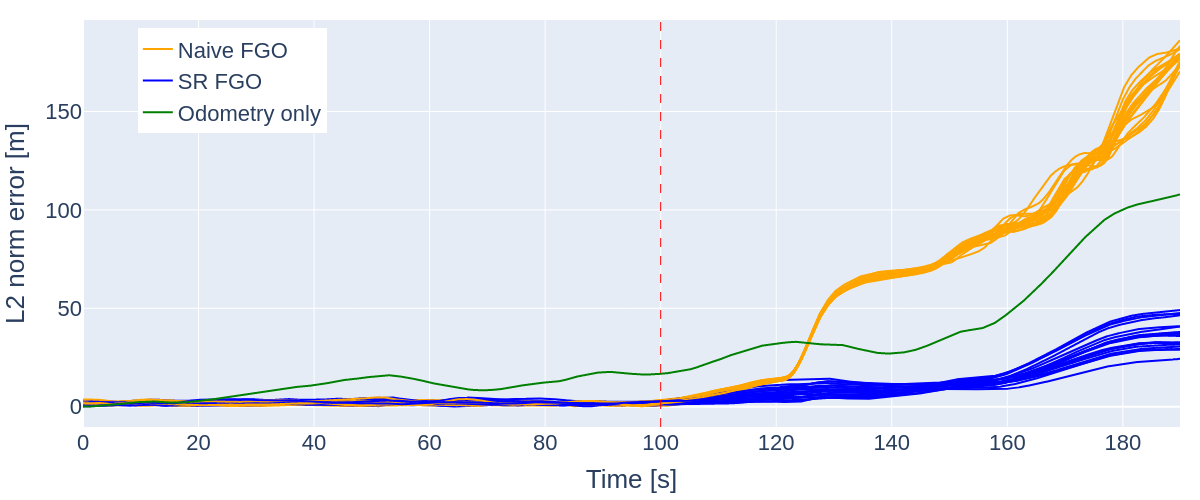}
         \caption{Sequence 0018}
         \label{fig:0018_errors}
     \end{subfigure}
     \hfill
     \begin{subfigure}[b]{0.49\textwidth}
         \centering
         \includegraphics[width=\textwidth]{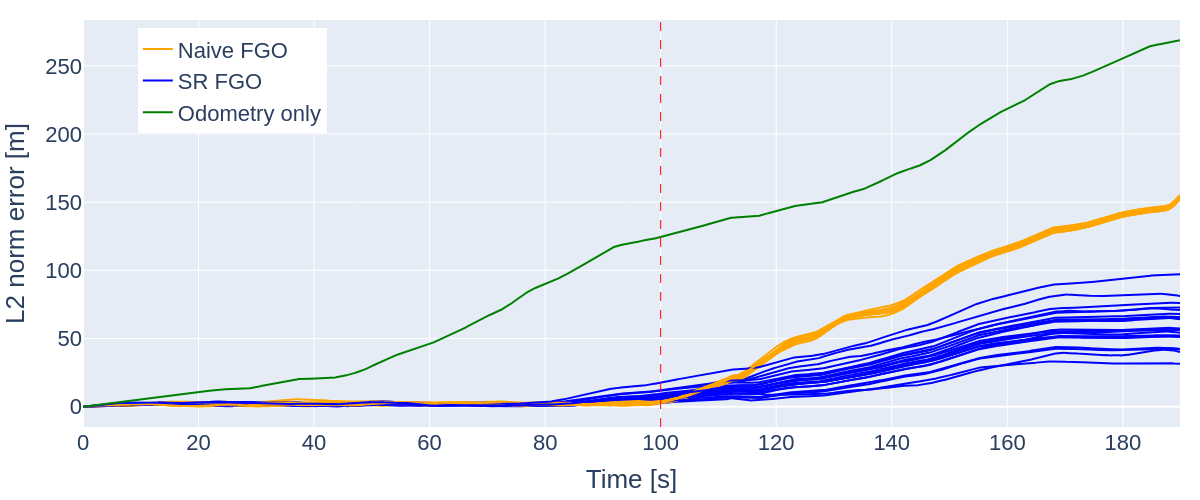}
         \caption{Sequence 0027}
         \label{fig:0027_errors}
     \end{subfigure}
     \hfill
     \begin{subfigure}[b]{0.49\textwidth}
         \centering
         \includegraphics[width=\textwidth]{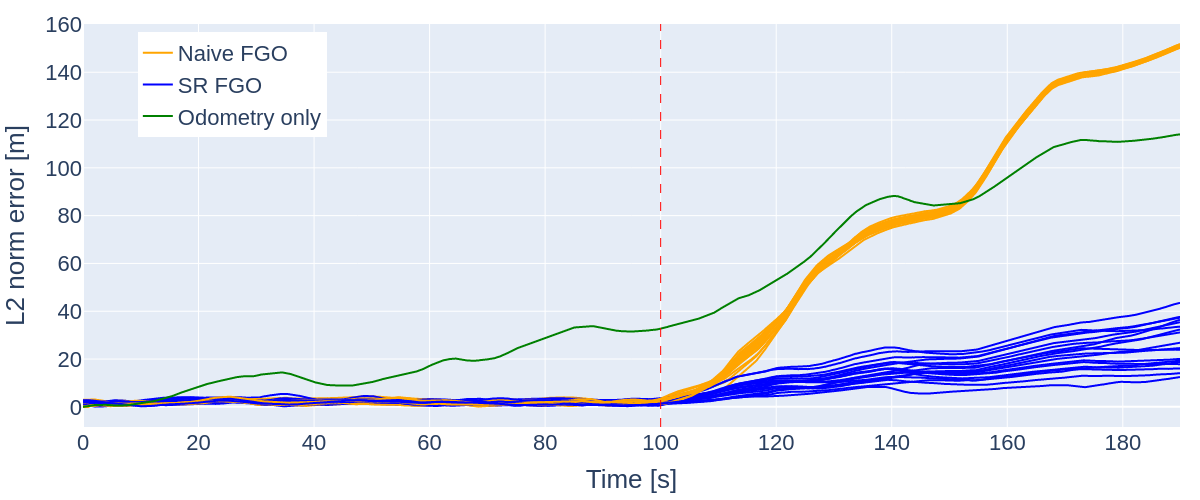}
         \caption{Sequence 0028}
         \label{fig:0028_errors}
     \end{subfigure}
     \hfill
     \begin{subfigure}[b]{0.49\textwidth}
         \centering
         \includegraphics[width=\textwidth]{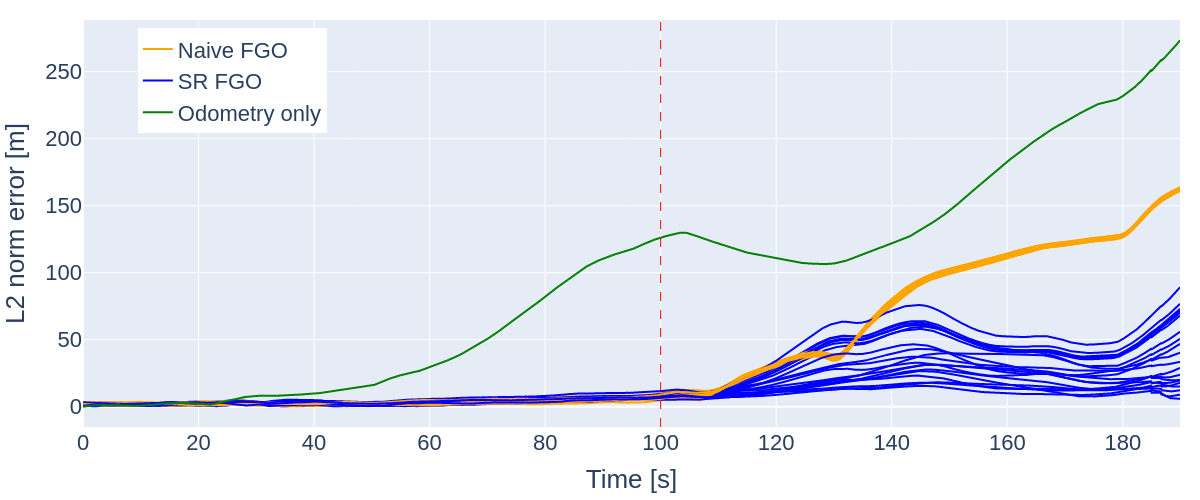}
         \caption{Sequence 0034}
         \label{fig:0034_errors}
     \end{subfigure}
    \caption{Comparison of trajectory errors for three different approaches on KITTI sequences. For each plot, LiDAR odometry error over time is shown in green, while 20 Monte Carlo runs each of Naive FGO and SR FGO are shown in red and blue respectively. In each sequence, our spoofing-resilient FGO mitigates the spoofing attack, bounding errors during the Chimera epoch to well under odometry drift levels. }
    \label{fig:error_comparison}
\end{figure*}

Next, we focus on the spoofed case of $r\spoof = 2.0$ \si{m/s}, and compare the $L^2$ position error over time of our SR FGO approach against the two baselines Odometry only and Naive FGO.
Fig.~\ref{fig:error_comparison} shows a comparison plot for each KITTI trajectory.
In each plot, 20 Monte Carlo runs of Naive and SR FGO are shown, and the start of the spoofing attack at 100 seconds is indicated by the vertical red dashed line.
For each trajectory, we see that LiDAR odometry suffers from significant drift over time, on the order of 100s to 200s of meters of final $L^2$ norm position error after 200 seconds.
For both Naive FGO and SR FGO, $L^2$ position errors remain under 5.0 \si{m} for the first 100 seconds during authentic conditions.
However, after the start of the attack at 100 seconds, the Naive FGO approach is heavily influenced by the spoofing attack, and its $L^2$ norm position error diverges, with final error exceeding that of LiDAR odometry in sequences 0018 and 0028.
On the other hand, our SR FGO approach is able to consistently detect and mitigate the spoofing attack, and keep position errors bounded to under odometry drift levels.

% One may observe that for sequence 0027, the SR FGO error starts increasing slightly before the start of the attack at 100 seconds. 
% This is due to our re-processing step for mitigation, in which the window 

% Errors over time

% \subsection{Chimera Authentication}
% %% -- Show how Chimera authentication is used by plotting error/trajectory over time and showing jump back to 0 (small) error at 180s

\subsection{Window Size Comparison} \label{subsec:window_comp}
%% -- Compare performance across window sizes

Now, we perform a case study to analyze to effect of varying window size on the performance of our algorithm.
Table \ref{tab:window_comp} shows a comparison of mean and max $L^2$ norm position error as well as average iteration time across a range of increasing window sizes for each sequence. 

\begin{table*}[htbp]
\caption{Window Size Comparison under 0.2 \si{m/s} Spoofing averaged over 20 Monte Carlo runs. Performance initially improves with increasing window size then saturates between 5 \si{s} to 20 \si{s}, while average iteration time increases consistently with window size as expected. Window size is expressed in timesteps (of length 0.1 \si{s}).}
    \begin{subtable}[h]{0.45\textwidth}
        \begin{center}
        \begin{tabular}{|c|c c c|}
            \hline
            \textbf{Window size} & \textbf{Mean error} & \textbf{Max error} & \textbf{Avg. iteration time} \\
            \hline \hline
            20 & 26.1 \si{m} & 101.0 \si{m} & 0.144 \si{s} \\
            50 & 8.66 \si{m} & 41.1 \si{m} & 0.469 \si{s} \\
            100 & 9.75 \si{m} & 46.2 \si{m} & 1.25 \si{s} \\
            200 & 11.5 \si{m} & 59.5 \si{m} & 3.57 \si{s} \\
            300 & 13.9 \si{m} & 63.6 \si{m} & 6.69 \si{s} \\
            \hline
        \end{tabular}
        \vspace*{0.5mm}
        \caption{Sequence 0018}
        \label{tab:0018_window}
        \end{center}
    \end{subtable}
    \hfill
    \begin{subtable}[h]{0.45\textwidth}
        \begin{center}
        \begin{tabular}{|c|c c c|}
            \hline
            \textbf{Window size} & \textbf{Mean error} & \textbf{Max error} & \textbf{Avg. iteration time} \\
            \hline \hline
            20 & 189.4 \si{m} & 372.3 \si{m} & 0.130 \si{s} \\
            50 & 28.1 \si{m} & 78.4 \si{m} & 0.445 \si{s} \\
            100 & 20.9 \si{m} & 63.9 \si{m} & 1.13 \si{s} \\
            200 & 29.6 \si{m} & 92.3 \si{m} & 3.37 \si{s} \\
            300 & 30.7 \si{m} & 94.8 \si{m} & 5.78 \si{s} \\
            \hline
        \end{tabular}
        \vspace*{0.5mm}
        \caption{Sequence 0027}
        \label{tab:0027_window}
        \end{center}
    \end{subtable}
    \hfill
    \newline
    \vspace{2mm}
    \newline
    \begin{subtable}[h]{0.45\textwidth}
        \begin{center}
        \begin{tabular}{|c|c c c|}
            \hline
            \textbf{Window size} & \textbf{Mean error} & \textbf{Max error} & \textbf{Avg. iteration time} \\
            \hline \hline
            20 & 47.6 \si{m} & 112.6 \si{m} & 0.139 \si{s} \\
            50 & 15.2 \si{m} & 44.3 \si{m} & 0.456 \si{s} \\
            100 & 12.5 \si{m} & 34.0 \si{m} & 1.21 \si{s} \\
            200 & 7.07 \si{m} & 24.1 \si{m} & 3.41 \si{s} \\
            300 & 15.9 \si{m} & 42.8 \si{m} & 6.47 \si{s} \\
            \hline
        \end{tabular}
        \vspace*{0.5mm}
        \caption{Sequence 0028}
        \label{tab:0028_window}
        \end{center}
    \end{subtable}
    \hfill
    \begin{subtable}[h]{0.45\textwidth}
        \begin{center}
        \begin{tabular}{|c|c c c|}
            \hline
            \textbf{Window size} & \textbf{Mean error} & \textbf{Max error} & \textbf{Avg. iteration time} \\
            \hline \hline
            20 & 83.8 \si{m} & 241.5 \si{m} & 0.144 \si{s} \\
            50 & 25.8 \si{m} & 110.4 \si{m} & 0.469 \si{s} \\
            100 & 23.8 \si{m} & 106.4 \si{m} & 2.64 \si{s} \\
            200 & 28.8 \si{m} & 127.3 \si{m} & 3.38 \si{s} \\
            300 & 25.5 \si{m} & 121.1 \si{m} & 6.43 \si{s} \\
            \hline
        \end{tabular}
        \vspace*{0.5mm}
        \caption{Sequence 0034}
        \label{tab:0034_window}
        \end{center}
    \end{subtable}
    \label{tab:window_comp}
\end{table*}

As expected, for all sequences, the average iteration time increases with window size, as the factor graph optimization must be done over a larger window with more measurements.
We also observe high rate of false detection for the smallest window size of 20.
This is to be expected, as the test statistic will be more sensitive to measurement noise and small errors in the trajectory estimates for a smaller window, and thus this window size behaves similarly to LiDAR odometry only in performance.

For all sequences, we also notice that improvement in mean and max error saturates as we increase the window size, occurring at $N = 50$ for sequence 0018, $N = 100$ for sequence 0027, $N = 200$ for sequence 0028, and $N = 100$ for sequence 0034.
This is most likely due to that fact that, as window size increases, a larger window of the trajectory is re-processed when spoofing is detected.
If spoofing is detected for a window with majority authentic measurements but some spoofed measurements towards the end, then we may discard more authentic measurements which may adversely affect the overall positioning performance.
The results of this case study validate our general choice of window size $N = 100$ for our experiments.

% For sequence 0018, we notice that performance improvement quickly saturates at $N=100$.
% However, for sequence 0027, performance stays relatively poor for window sizes of $N = 100$ and 200, then starts improving at $N = 300$ and 400. 
% We believe this is due to the nature of the trajectory, in which a large spoofing attack such as $r\spoof = $ 0.2 \si{m/s} immediately induces significant heading error which combined with LiDAR odometry drift results high errors.
% This trajectory benefits from the larger window sizes as they are able to see the attack coming from farther away, and also upon detection re-process a larger window for mitigation, and thus more effectively removing the effects of spoofing from the trajectory estimate.

\subsection{Detection Statistics}
%% -- Test statistic monte carlo plots, detection rates, false alarm rates

\begin{figure*}
     \centering
     \begin{subfigure}[b]{0.49\textwidth}
         \centering
         \includegraphics[width=\textwidth]{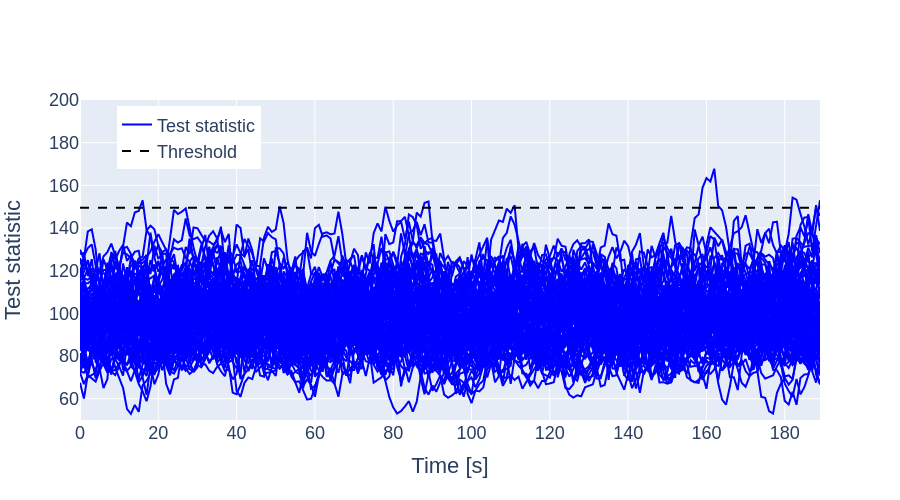}
         \caption{Spoofing test statistic plotted for 100 Monte Carlo runs under nominal (authentic) conditions. Our detector satisfies the prescribed false alarm rate of $\alpha = 0.001$.}
         \label{fig:test_stat_nom}
     \end{subfigure}
     \hfill
     \begin{subfigure}[b]{0.49\textwidth}
         \centering
         \includegraphics[width=\textwidth]{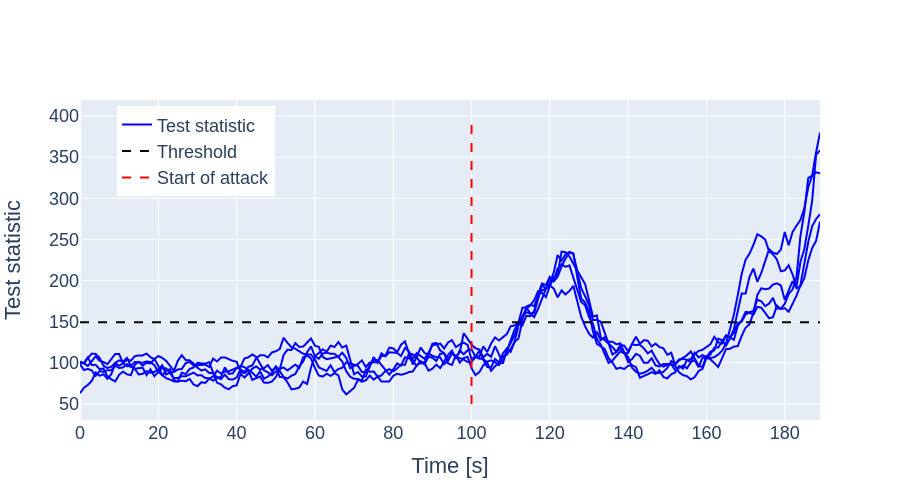}
         \caption{Spoofing test statistic plotted for 10 Monte Carlo runs under a spoofing attack of $r\spoof = 1.0$ \si{m/s}. The statistic crosses the threshold shortly after the attack for all 10 of the runs. }
         \label{fig:test_stat_spoof}
     \end{subfigure}
    \caption{Monte Carlo simulations of spoofing test statistic plotted over time for both nominal and spoofed conditions. Our detection method remains under the user specified false alarm rates, while consistently detecting the presence of a sufficiently large spoofing attack.}
    \label{fig:detection_stats}
\end{figure*}

Finally, we examine the detection statistics for our algorithm, first running 100 Monte Carlo simulations for the nominal, unspoofed case to test the false alarm rate of our detector. These runs are plotted along with the corresponding detection threshold in Fig.~\ref{fig:test_stat_nom}.
The parameter $\alpha = 0.001$ corresponds to the false alarm probability of a single trial, so across a 180 second long Chimera epoch with 180 trials (one for every GPS measurement at 1 Hz), the probability of a false alarm occurring during the Chimera epoch is $1 - (1 - \alpha)^{180} = 0.165$. 
Across the 100 Monte Carlo runs, there were a total of 9 false alarms, for an empirical per run false alarm rate of 0.09, and 15 total individual trial false alarms out of 18000 individual trials for an empirical per trial false alarm rate of 0.000833.
Thus, we see that our detector satisfies the desired false alarm rate requirements set by the user.

% MC test statistic plots (N = 10)

We also perform Monte Carlo simulation for the spoofed case for an attack with $r\spoof = 1.0$ \si{m/s}, shown in Fig.~\ref{fig:test_stat_spoof}.
In each of the 10 runs, the test statistic crosses the threshold shortly after the start of the attack, successfully detecting it, with a average time to detect of 11.2 seconds.

\section{Conclusion} \label{sec:conclusion}

In this work, we present a new framework for spoofing-resilient LiDAR-GPS factor graph fusion for Chimera GPS, which provides continuous and secure state estimation between Chimera authentication times. 
Our approach fuses LiDAR and GPS measurements with factor graph optimization, and computes a test statistic for spoofing detection based on the GPS factor residuals.
From this test statistic, our approach determines when to leverage the unauthenticated GPS measurements during the Chimera epoch, in order to improve localization performance when GPS is likely authentic.
We evaluate our approach with real-world data from the KITTI self-driving dataset, using sequences which span the Chimera slow channel 3-minute epoch. 
Our results demonstrate rapid detection and effective mitigation of spoofing attacks during vulnerable periods between authentications.

%This problem is challenging as LiDAR odometry alone suffers from significant drift, 

% More commentary on results
This work contributes towards the problem of designing LiDAR-GPS factor graph localization that is robust to GPS spoofing attacks. Our approach is designed around the Chimera signal enhancement, which will be a critical utility to authenticating GPS measurements against spoofing. Between Chimera authentications, we utilize the LiDAR sensor measurements to validate and strategically leverage GPS measurements to improve localization performance during authentic conditions, while maintaining resilience against experienced attacks during spoofing. Our work addresses the research gap for LiDAR-GPS fusion platforms, and takes an important step towards ensuring continuous navigation security for users of the future Chimera-enhanced GPS.

\FloatBarrier

\section*{Acknowledgment}

This material is based upon work supported by the Air Force Research Lab (AFRL) under grant number FA9453-20-1-0002.
We would like to thank the AFRL for their support of this research. 
We would also like to thank Shubh Gupta for reviewing this paper.

\bibliographystyle{IEEEtran}
\bibliography{references}

\end{document}